# Intersectional Inquiry, on the Ground and in the Algorithm

Shanthi Robertson[1], Liam Magee[1] 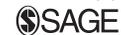, and Karen Soldatić[1]

## Abstract
This article makes two key contributions to methodological debates in automation research. First, we argue for and demonstrate how methods in this field must account for intersections of social difference, such as race, class, ethnicity, culture, and disability, in more nuanced ways. Second, we consider the complexities of bringing together computational and qualitative methods in an intersectional methodological approach while also arguing that in their respective subjects (machines and human subjects) and conceptual scope they enable a specific dialogue on intersectionality and automation to be articulated. We draw on field reflections from a project that combines an analysis of intersectional bias in language models with findings from a community workshop on the frustrations and aspirations produced through engagement with everyday artificial intelligence (AI)–driven technologies in the context of care.

## Keywords
automation, intersectional research, disability studies, focus groups, methodologies

## Introduction

This article makes two key contributions to methodological debates in automation research. First, we argue for and demonstrate how methods in this research space must account for intersections of social difference (such as race, class, ethnicity, culture, and disability) in more nuanced ways. Second, we consider the complexities of bringing together computational and qualitative methods in an intersectional methodological approach while also arguing that in their respective subjects (machines and human subjects) and conceptual scope they enable a specific dialogue on intersectionality and automation to be articulated. We draw on field reflections from a project that combines a partly automated analysis of intersectional bias in language models with qualitative analysis from community workshops, interviews, and focus groups on how intersections of disability, race, class, and culture impact people's engagements with everyday artificial intelligence (AI)–driven technologies.

We critically reflect on intersectional research praxis at two levels: the intersection of methods and intersectional identities within the communities we research with. Engaging illustrative research encounters in our work so far, we consider: how do we operationalize interdisciplinary methods in the study of automation and inequality with attentiveness to how intersectional identities are both lived "on the ground" and processed through algorithms? As scholars across these fields, how do we develop a shared language to interrogate and understand autonomous technologies between disciplines and with technology and community stakeholders, advocates and community members? Moreover, how do we understand how power is replicated or transformed through this language and our research encounters?

Our methodological arguments build from the premise that users at the intersection of ethnic and cultural differences and disability identities remain notably excluded from advances in automated technology developments and innovations as market and design principles for everyday consumer technologies assume nondisabled and culturally homogeneous users. Such bias is even more marked with the rise in AI systems trained on datasets with racist and ableist skew (Noble, 2018; Pasquale, 2015).

Scholarship has identified how the quotidian experiences of users living at the intersection of culturally and linguistically diverse and disability identities can trouble notions of technology's social inclusion functions and problematize the assumptions that underpin them. Minority

[1]Western Sydney University, Penrith, New South Wales, Australia

**Corresponding Author:**
Liam Magee, Western Sydney University, Locked Bag 1797, Penrith, New South Wales 2751, Australia.
Email: l.magee@westernsydney.edu.au

users remain marked by racial, ethnic, and cultural differences despite technology's "assistive" capacities (Parette & Scherer, 2004). Assistive technologies can re-enforce ethnocultural assumptions in relation to the pathologization of disability and impairment as an individualized deficiency, further reproducing social–cultural relational structures of disability stigmatization and marginalization (Parette & Scherer, 2004). Yet autonomous technologies also have a range of potential positive impacts on individual users and communities living at the intersection of cultural, racial or ethnic, and disability identities. Everyday consumer technologies designed for mainstream markets such as smartphone AI companions, home speakers, or translation and speech synthesis applications are already in frequent use in these communities to enable everyday tasks associated with health, leisure, and communication (Locke et al., 2021). As examples, Google Translate is used frequently in non-English background communities in Australia to manage social and service interactions. Voice-to-text applications and automated captioning are similarly used for communication by the Deaf community and people with hearing impairments (Morris et al., 2014). Home speakers are used for communication, reminders, and entertainment by people with physical disabilities and dementia in their home environments (Bier et al., 2018). And carers, support workers, and people with disabilities and/or chronic conditions deploy various consumer technologies to coordinate and enhance everyday support routines (Guchet, 2021).

We reflect upon methodological approaches from the first phases of a wider project, the *Autonomy, Disability and Diversity: Everyday Practices of Technology (ADDEPT) Project*, which examines AI technologies and their current and potential impacts on people living with disabilities from culturally and linguistically diverse backgrounds in Sydney's southwest and northeast. Methodologically, the project takes an engaged, participatory, and primarily qualitative approach, working in collaborative partnership with local multicultural and disability organizations, community arts and culture spaces, and technology companies to engage people with disability from diverse cultural and ethnolinguistic minority backgrounds in the current and potential use of AI-driven technologies through a series of activities involving self-documentation of technology use and creative experimentation with new and emerging consumer technologies. As an interdisciplinary team across disability studies, sociology, and computer science, and as described further below, we have also collaborated on experiments testing the presence of intersectional bias in language models.

Conceptually, the project combines three strands of sociological research—disability, migration, and science and technology studies—into productive dialogue to generate new insights and knowledge within and between these fields. We argue the combination of qualitative social and quantitative computational methods produces new understandings of human–machine interaction by examining how human subjects think AI through metaphors and how machines—specifically language models—in turn generate specific associations with human subjects, when modified by linguistic markers of social difference. We build on studies of technology in relation to disability and cultural and ethnic diversity which acknowledge the intersectional, relational, adaptive and embodied nature of technology use (e.g., Ellis & Kent, 2011, 2017; Moser, 2006; Söderström, 2009a, 2009b). Intersectionality (Collins, 2000; Crenshaw, 1989; Combahee River Collective, 1977/2014) underpins the project's design, conceptually and methodologically, as we elaborate, both empirically and theoretically, on the relationship between embodied and intersectional social difference, social inclusion, and autonomous technology. Intersectionality grounds an analysis and critique of how AI in everyday life has too often been limited to mainstream populations or to singular categories of social difference. Research has long recognized that technology design and user studies have been poorly applied to minority populations (Alper, 2017; Blume, 2012; Introna & Wood, 2004; Johnson & Moxon, 1998) and rarely extended to groups at the intersections or margins of common categorical distinctions, such as a primary language group or specific disability type (e.g., physical, sensory, cognitive, intellectual, and/or neurological). While we acknowledge the significance of the core activist practices of intersectional thinking, following Gillborn (2015) we position intersectionality in this article primarily as a tool of analysis and center the empirical claims of Crenshaw and other theorists in seeing intersectional approaches as critical methodologically to understand more fully how social inequities are created, reproduced and sustained.

Here we focus on two methodological strands of this wider project: qualitative analysis of community scoping workshops that explore how intersections of disability, race, ethnicity, class, and culture impact on people's engagements with everyday AI-driven technologies, and a computational and partly automated analysis of intersectional bias in language models. These two strands act as distinct but interrelated encounters between academic disciplines, between methodological approaches, and between participants and researchers that highlight the challenges and possibilities of researching "everyday" manifestations of AI through the lens of intersectionality. Centering the sometimes unresolved tensions of the research encounter, we reflect on two ongoing concerns that continue to inform the methodological and analytic possibilities of this work: how the descriptive politics of AI manifest through and shape the fieldwork in relation to language, imagery, and metaphor and how attunement to intersectionality generates new understandings of misreading and misrecognition both on the ground and in the algorithm.

## Automation Metaphors and Algorithmic Bias

As researchers gathering data on everyday perceptions, experiences, and effects surrounding consumer AI technologies, we understand our research encounters as moments through which visions and understandings of AI are mediated and produced. As Ziewitz (2016) noted, computational artifacts like algorithms or AI systems can function as sensitizing devices that can challenge entrenched assumptions, including assumptions about agency, autonomy, and normativity. Here we mobilize the research encounter, between researchers and participants, and between researchers and data, as moments through which to unfold and reflect on the assumptions that underpin research design and methodology.

These moments are embedded within broader public, technocratic, and regulatory discourses on AI, yet are also encounters through which such discourses are potentially reconfigured or destabilized or in which more marginal or variously culturally situated visions of AI come into tension with each other. They speak to how, in our fieldwork, we grapple with the ways AI is culturally and interculturally "talked into being" (Bareis & Katzenbach, 2021) in the moment of the research encounter. We thus build on the cultural and social analysis of AI that has foregrounded the constitutive work of metaphor, imagery, and fictionality (Campolo & Crawford, 2020; Corbyn, 2021; Esteban Casañas, 2020). Metaphors such as "intelligence" or "learning" "guide the societal discourse sustainably and fuel fantasies and future visions in the broader public just as much as in expert communities" (Bareis & Katzenbach, 2021, p. 3). Metaphors and language demonstrably shape not only technocratic and public discourses but also scholarly debates in ways that overlap with established disciplinary norms within the study of technology. As Lupton (2014) points out, for example, the long-standing focus in cultural studies on the "cyber" emphasizes futuristic, science fiction metaphors. In contrast, sociological or pedagogical literature that refers to "information technologies" draws more attention to grounded and utilitarian questions of use and access. Yet everyday perceptions of AI in the context of support and/or care for people with disabilities from culturally and linguistically diverse backgrounds remain notably understudied (Yigitcanlar et al., 2020). With few exceptions, analysis of how AI is "talked into being" remains focused on the bureaucratic or technocratic discourse, or on broad definitions of "public perception" that pay little attention to how intersectional social difference, social marginalization, or culture may shade or shape these perceptions. In contrast to research on mitigating algorithmic bias, which has begun to engage with intersectionality (Agrawal, 2019; Buolamwini & Gebru, 2018; Kim et al., 2020), the discussion of the discursive politics of AI is yet to pay much attention to how intersections of disability, race, ethnicity, gender, culture, class, generation, or place might shape individual and collective imaginaries. Further to this, we use reflections from our fieldwork to focus on how the discursive politics of AI concerning intersectional social difference shape interdisciplinary research encounters and methodologies in the investigation of Corbyn & Crawford, 2021 "everyday" understandings and use of technology.

Our second encounter examines how technology in turn imagines intersectional social difference through the ways language models predict sentences following on from input that include one or more markers of such difference. We focus on *bias*: a topic that motivates a now wide computer science literature that seeks to address gender, racial, and other biases within these models (Bender et al., 2021; Bolukbasi et al., 2016; Bordia & Bowman, 2019), with limited but growing attention to the interaction between these biases (Buolamwini & Gebru, 2018; Guo & Caliskan, 2021; Kim et al., 2020). We treat bias in ways consistent with some of that literature (e.g., Dwork et al., 2012) in a narrow and specific sense: we submit sentences produced by language models to a separate automated sentiment analysis system, which produces a measure of negative-to-positive sentiment (between 0.0 and 1.0) based on how that second system evaluates those sentences. We determine bias by whether a collection of such sentences prompted by one set of social terms scores significantly higher or lower on average sentiment scores. For example, if a series of sentences generated by a language model in response to prompts that include the word "Muslim" score substantially lower or higher than other markers of religion (such as "Hindu" or "Jewish")—or indeed the absence of any such terms—we determine that language model biased.

While consistent with other computational studies, this approach to bias raises other questions. AI language systems are after all trained and evaluated on their ability to associate certain words with others, to produce a sensible and human-understandable response to questions, search engine queries, and so on. In this sense, language models are from one point of view essentially *engines of metaphor*—they are trained to "carry across" associations gleaned from iterated analysis of large text corpora into new linguistic productions. One question relates to the use of sentiment analysis, itself an example of a language model that judges, rather than generates, sentences and other language use. Sentiment analysis can import bias of its own, attribute positive or negative sentiment to utterances that human readers would disagree with. But of interest here are questions that instead connect the narrow technical meaning of bias to its broader social causes and effects.

Wellner and Rothman (2020) discuss the complex causality of bias in ways helpful for connecting the narrowness of the treatment of bias in our second study to the more open considerations of metaphor in our first. They consider four approaches to addressing bias: examining and curating data sets prior to system training; algorithmic transparency, so the (at least technical) causes of bias can be inspected; *ex post* bias identification and mitigation by other "supervising" algorithms; and user review and oversight, particularly by those subject to discrimination as a result of bias. The range of these approaches, across stages of system design and use and involving multiple sites of agency, highlight the fact that de-biasing AI systems is always a more-than-technical activity, and in positioning a study with a narrow interpretation of bias against a broader consideration of the sociality of AI, we draw attention to how automated decisions are enacted by multiple sites of agency and are determined by data-driven images and metaphors of distinct social positions. In terms of their effects, embedded and threaded into the infrastructure of cloud-based services evoked by everyday users, these decisions are also decisive normatively and politically, determining symbolic relations that echo, perturb, erase, or amplify those experienced in other social fields. For people who identify or are identified with intersectional categories, these determinations can reinforce existing systems of social exclusion and discrimination.

## Understanding the Care Robot: The Descriptive Politics of AI Metaphors in the Qualitative Research Encounter

We draw upon a vignette from our fieldwork as our first reflective case. Two project researchers, Snow and Shanthi, and a student researcher on placement, Julia, conducted a focus group workshop in one of the project's target regions in Sydney with parent carers who were first-generation migrants from East Asian backgrounds. The group consisted of 17 carers in their 50s and 60s who have been members of a local parental carers' support group for about 7 years. Participants had migrated to Australia from China, Hong Kong, Malaysia, Korea, and Japan. Most participants had functional to fluent English, but the workshop was conducted in multiple languages, with Snow interpreting in Mandarin and Cantonese and Julia in Korean when needed. Ashley, another student researcher fluent in Mandarin and Cantonese, provided further translation when transcribing the data. The participants were all full-time parental carers of their adult children who were on the autism spectrum. Their adult children were all defined as having high support needs, although there were various individual support needs within the group.

During the first half of the focus group workshop, we discussed the technologies that parental carers and the persons with disabilities that they support used every day in their routines. The participants used a wide range of technologies, some powered by AI, from expensive assistive devices for communication to mainstream smartphone applications, like white noise apps or the iPhone "memories" function, to facilitate communication, well-being, or desired behavioral outcomes. The parental carer participants expressed several concerns about data, privacy, and safety, especially when discussing communication apps and social media. In the final half of the workshop, we asked participants, working in small groups, to speculate on the future of AI and to design a technology that they might find useful in their daily lives as parents that provide ongoing daily support to their adult children living with disability in the family home. In previous focus groups, which we conducted with mostly White service providers and disability service providers, fears and anxieties around AI "taking over" the workforce and removing the "human element" from disability support work had dominated the discussions of AI futures. So facilitators were prepared for similar anxieties to be raised, especially given the existing concerns discussed in the morning sessions, the age of participants, and the fact that, despite being regular users of technology, most participants had self-identified in these discussions as having limited capacities and understandings around new technologies due to their age as well as language and cultural factors.

In this section, we draw on the first-person notes of one of the researchers and facilitators (Shanthi) as well as direct quotes from the workshop's transcript to illustrate aspects of this workshop not only as qualitative data but as a research encounter within which the descriptive politics of AI mediates what data is collected, and how different visions of AI are entangled across multiple layers of interpretation and translation:

> "Do you know Doraemon?" H. (Japanese, male, 50s) asks me as he holds up his group's design brainstorm on butcher's paper. I immediately visually recognise the smiling blue manga character that H. has carefully sketched in the corner of the paper, but I am not familiar with its name or any other context. The character is, as H. partially explains, a robot cat. H.'s group has envisioned a life-size Doraemon care robot that H. shares will conduct multiple practices of caring:
> 
> So it may be a bit futuristic, but in Japan, everyone was growing up . . . kind of looking at Doraemon. Everyone knows that Doraemon. Everyone wants a Doraemon in the future. So for, for us, because in my case, my daughter, all these house security cases, safety problems, demolished things in danger, they may see she cut her fingers or everything. So some care robot reminder her of security, safety. You don't do that dangerous actions. And also my daughter has continence problems . . . And so, so first they stop it, and still she does it, just clean the room before school, and all this teaching new languages and like a human rules, the importance of consideration to other people, et cetera, et cetera . . . The Doraemon will be an interpreter because she, he, is an AI, and he understand her or him. He's a special terminology. Did they all speak language in a different way from us? They have to know what exactly means by those words, implications and translate to other carers, for us. And anyway, so all the years they moved more and more, a better companion, like a very good friends and reliable companions over years and growing together and sharing all that experience together. And when she is said, just comfort and console and making her happy, [people laugh], something like that, a futuristic, any robot like this, they're happier.

As we go around the groups in the room, each group has designed a version of a care robot. Some are animals, some are more "android-like," but all, like the Doraemon described by H.'s group, are effectively replicants of a human carer, but with advanced powers like strength and communication. The functions noted by H's group are often repeated: keeping care recipients safe from accidents in the home, cleaning up from incontinence or other mess, behavioural control. Communication across both linguistic and cognitive difference is also central in these discussions. Because some care recipients have limited verbal capacities, and all services and many technologies are only provided in English, all the parent-carers communicate with their children in English, even when their own English is limited. Multiple layers of translation across linguistic and communicative difference are thus central to many of the group's imaginaries of the care robot. H. and his group, for example, see their care Doraemon as able to bridge the communication gap between carers and care recipients, to "interpret" beyond the limits of human communication across cognitive difference. Eventually, over time, the Doraemon robot will become a friend who can comfort and console and provide emotional support. Other groups similarly talk about their care robots being able to "hug," to "play" and to "dance" with both carers and care recipients.

I was initially somewhat taken aback by these utopian visions of "cute," cartoon-like care robots dominating every group's design. Later, Ashley provided a deeper explanation of Doraemon and his cultural significance in her transcription notes, and of course, I turned to Google for further context. While still popular with young people today, I learned that Doraemon was first serialised in Japan in 1969. Unlike the androids and cyborgs familiar in the cultural scripts of my own childhood, who were mostly either servile or threatening, Doraemon, a cat robot, helps a young boy Nobita with childhood problems like schoolwork and bullying. Doraemon has been sent from the future by Nobita's grandson to guide his grandfather through childhood so his adulthood, and subsequently his children and grandchildren's lives, can be enhanced. As Doraemon helps Nobita overcome his troubles, they become close friends. Although produced and set in Japan, Doraemon was broadcast throughout Asia in the 1980s and 1990s, and only broadcast in the US and Australian after 2013. Although incredibly popular and instantly recognisable through most of East and Southeast Asia, Doraemon has never gained the same popularity in the West as other Japanese animations like Dragon Ball Z or Pokemon. However it shares with these other icons of Japanese mecha anime culture a focus on cute pet-like or child-like robots that at the same time extend care and protection towards their owners or companions—a mix of features that as Allison (2006) has documented, proved commercially successful in US and other markets without losing an air of subcultural chic and Japanese exoticism.

Many scholars have noted how popular discourses of AI continue to be grounded in motifs drawn from science fiction and other popular culture representations of anthropomorphic machines (Natale & Ballatore, 2017). The discussion of care robots in this focus group was a moment that required some translation between researchers and participants for different cultural visions and embedded understandings to be mutually recognized. Doraemon and his attached meanings—as a robot who is both a friend and a child's carer—had instant cultural recognition between the participants and Ashley during transcription. Yet the pop culture knowledge of researchers raised in Australia did not overlap with these shared understandings of the character and its meanings.

An intersectional lens is generative in understanding this research encounter and in analyzing the data it produces because the visions of an AI future that these carers created together were inflected by both their own cultural and generational identities and their specific experiences as carers who provide daily support to their adult children with disabilities within the family home. Rather than creating assistive or curative visions of technology that were designed to mediate or cure impairments for people with disability or bring more independence or autonomy for persons with disabilities in relationship with carers, workshop participants uniformly envisioned care robots that would effectively replicate and amplify the role of a human parental carer, providing practical embodied care, emotional nurturing, and behavioral surveillance and control (see Fig. 1). The robots were also imagined as translators, enabling communication at the intersections of linguistic and neurological difference.

This vision of an AI-driven disability-care future is only utopian from the carers' perspective and not necessarily from the perspective of persons with disabilities who are subject to long-standing familial relations of care. Furthermore, the dominance of the figure of the care robot in this particular research space rendered less visible other more probable visions of AI and disability. Our collective knowledge, as researchers of AI and the social, of likely technological futures that may impact these participants, their adult children, and their communities remain unspoken in the mode of the focus group method, in which, despite acknowledgment of the co-construction of knowledge, the "collection" of participant perspectives retains primacy. The "care" robots that are already beginning to infiltrate disability and aged care are not human-sized "friends" who replicate, with enhanced capabilities, the practical and emotional care of family or professional human support workers. Instead, they are codes, apps, and algorithms that indeed perform some of the everyday support work the parental carers describe for their adult children (such as reminders about medication or behavioral reinforcement) but also monitor and surveil people with disability and the care and support workforce (e.g., Panico et al., 2020; Stamboliev, 2020).

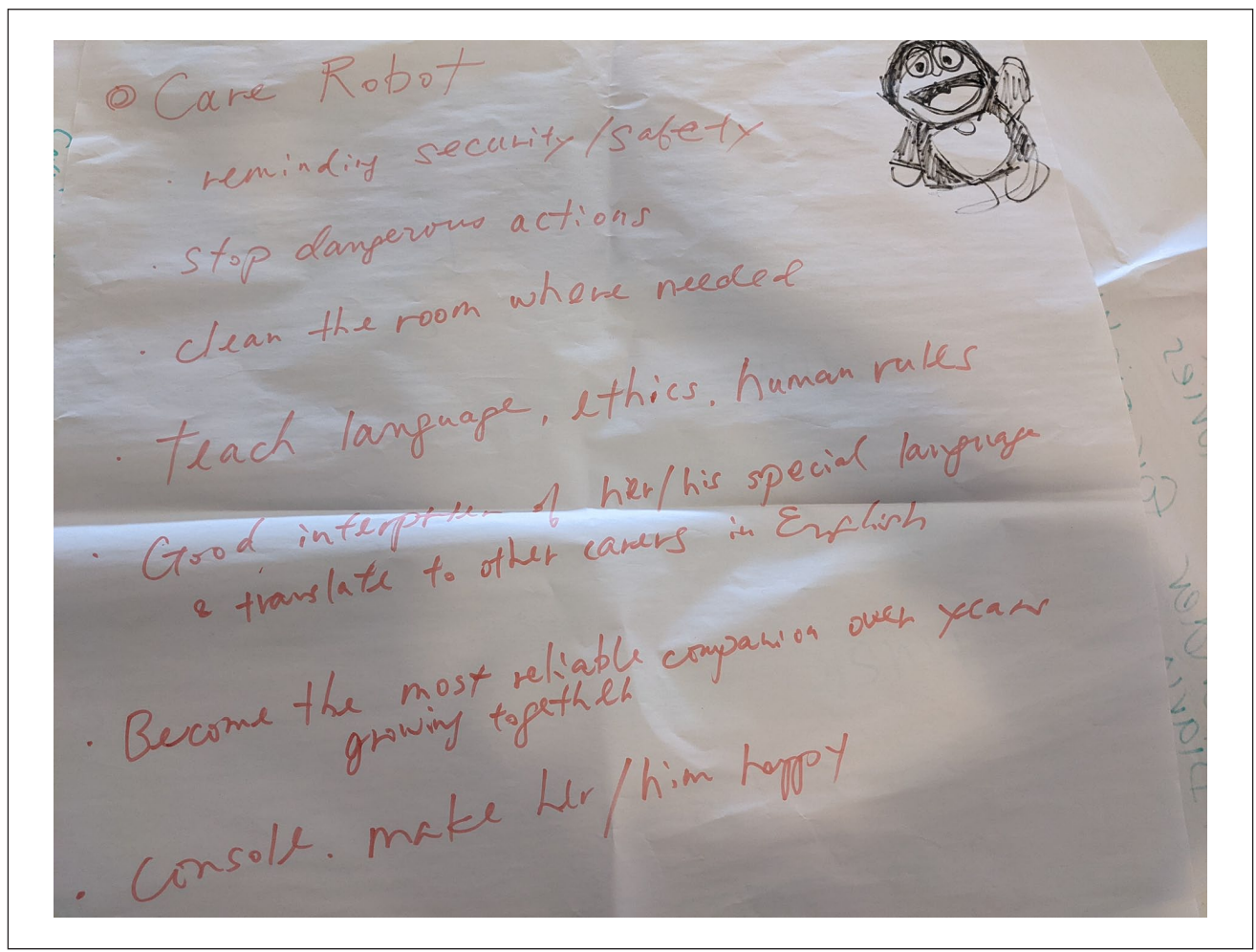

**Figure 1.** The design brainstorm from H's group featuring H's sketch of Doraemon.

These versions of the care robot are unlikely to be designed with the complexities of our research participants' cultural or linguistic contexts in mind. Nor will they operate in alignment with the cultural vision of the robot produced in our workshop activity, in which robots are autonomous closed technological systems, whose "learning" and "belonging" remains bounded within the private domain of the household or family, in the manner of a family pet. Instead, the incipient future of the care robot involves interconnected technological architectures within which data collection and new forms of surveillance of domestic space (Lafontaine, 2020) will form a key part of value extraction for technology companies and potentially also for government actors. As researchers within the field of Human-Centered AI have rightly argued, there is a need to "shift language, imagery and metaphors away from portrayals of intelligent autonomous teammates towards descriptions of powerful tool-like appliances and tele-operated devices"

(Shneiderman, 2020, p. 109). In the context of our fieldwork, while the co-construction of meaning and visions of AI with our research participants is central to understanding how such meanings are locally and culturally situated, the "cute" cultural vision of Doraemon perhaps intercedes in the research encounter as both a moment critical to qualitative interpretation and an obstruction—a metaphor that blocks a different kind of dialogue within the research space of more probable futures.

Equally, we might ask whether the "appliances and tele-operated devices" that operate within care environments constitute blockages of their own, producing a certain sterile and clinical aesthetics of care that, no less than the "cute" AI envisioned by the carer group reflects a limited imagination on the part of their designers. As Markham (2021) has argued in relation to efforts to provoke novel public "imaginaries" of technology, existing cultural narratives and practices exert powerful—powerful but not "unbreakable"—influences over what is possible to imagine. This parental carer group expresses desires for technology to overcome barriers to communication and isolation they see that those they care for—and care about—face. Such expressions see technology as neither curative nor anthropomorphic substitutes but rather as agents that can overcome other limits: of an ableist society itself incapable of moving beyond normative constraints to provide companionship to people who require complex relational care.

## Human–Machine Intersections

Our second fieldwork vignette involves a very different field and site of encounter with AI. It is however no less linked to questions of communication, vision, and metaphor. While the first vignette discussed metaphor in the context of human projections and aspirations for the machine, in this second example, we see how the trained machine itself imagines diverse human subjects. To contextualize this, we need to explain that one major field of AI is in the understanding and interpreting of natural language, an area with important applications and implications for people who might benefit from automated machine translation (for nondominant language speakers), text-to-speech rendering (for vision-impaired people), auto-captioning (for deaf or hard-of-hearing people), and rewriting of texts into plain or easy English (for certain people with neurodiverse conditions). We were concerned with how AI systems themselves are beginning to "imagine" intersectional categories of disability and ethnic, racial, and cultural difference. We joined with a partner investigator at Microsoft to conduct an evaluative study of intersectional bias in causal language models, examining two models which could be freely accessed and evaluated on commodity hardware GPT-2 (developed and released by OpenAI in 2019, and since succeeded by GPT-3) and GPT-NEO (an open-source variant of GPT-3, released in early 2021).

Our experiments involved creating a limited number of terms representing disability, religion, and gender. We then developed an exhaustive list of prompts that included all possible 280 combinations of one, two, or three of these terms and were inputted to both language models (and all of their available size variants). The language models would then output the original prompt followed by a sequence of predicted words, which we converted into sentences. We used a separate sentiment analysis algorithm to judge whether these sentences could be interpreted as positive or negative. As discussed elsewhere (Magee et al., 2021), we found that compared with the lack of any marker, all gender markers, one marker of religion (Muslim) and almost all disability markers (especially physical), elicited lower average sentiment scores than others. This confirmed two tests of language models: (a) as previous literature has shown, they exhibit bias toward specific groups (e.g., people with physical disabilities, Muslims) and (b) this bias is also "intersectional" in narrow and technical sense, compounding as multiple terms are added to input prompts.

Our tests also examined differences in training data used by two language model systems (GPT-2 and GPT-NEO) and differences caused by training times (reflected in the parameter size of the models). In general, the more diverse training data used by GPT-NEO (as described by Gao et al., 2020, which references addressing bias through more diverse text inclusion) and the more diverse the associations (produced by longer training), the more positive scores overall—however, bias, as measured by differences in average scores between different categories, was not necessarily eliminated, and new examples of bias emerged. In several cases, we also found that intersectional bias could not be easily predicted by individual marker scores, which suggests that work on eliminating, for example, bias toward gender, disability, and religious categories is insufficient for addressing bias toward all possible combinations of these categories in language models.

To "reverse engineer" some of the associations between intersectional markers and predicted terms that led to positive or negative sentiment scores, we applied topic modeling to sentences generated by all eight models for high- and low-scoring prompts. The associated topics illustrate common stereotypical associations from media representations that the language models have varied ways internalized through training on content largely sourced from the internet. Terms associated with "a blind Muslim man"—one of the low scoring prompts—for example reference to violence and victimhood ("killed," "attacked," "accused," "beaten"), criminality ("state," "police," "arrested," "court," "accused"), and, more sparingly, religion and location ("Mosque," "Saudi," "Islamic"). Terms associated with "A Buddhist person with Down Syndrome"—a prompt registering comparatively high sentiment scores—instead reference persons and family ("child," "mother," "adult"), spiritual, psychological, and social states ("meditation," "experience," "belief," "depression," "different," "normal") and common terms associated with neurodiverse disability ("syndrome," "diagnosed," "treated," "condition").

Unlike earlier neural networks (such as *word2vec* and *gloVe*) that created direct links between individual words, recent language models like those examined here employ so-called "transformer" architectures (Vaswani et al., 2017) that bind words to other words through sentential context. Transformer-based systems are in a very literal sense then metaphorical machines or engines: they *carry across* associations of both individual and compound terms into their work of calculation and prediction. "Blind," "Muslim," and "man" draw together in a semantic space that, for these machines, is disconnected to any number of broad ways of being human—creative, political, scientific, historic, circumstantial or compassionate—and is instead bound up with perpetration or victimhood at the scene of crime or terror.

How the trained machine infers these associations can be redirected in many ways. Our experiments involved "zero-shot" training, where we supplied no prior examples of sentences we would like to generate, and research has shown language models are highly suggestive to techniques of fine-tuning and "few-shot" learning. We also used a single common set of parameters in generating outputs, and acknowledge different parameters can increase the diversity of associations (as well as the likelihood of nonsensical outputs). The approaches we adopted for detecting bias through sentiment analysis and modeling topics, while common in the algorithmic bias literature, present other variables to interpretation as well. Our aim in these experiments was not to reify machine associations as indelible, but point to ways that, when turned to practice, they could reproduce and amplify existing prejudice toward marginalized intersectional groups.

In engaging in the politics of computational language association, we arguably reproduce a common AI metaphor ourselves: of a biased and mistaken machine that could be corrected, even perfected. Even in designing the settings of our experiment—in the selection of gender, disability and religion as the intersectional categories of interest, and in the specification of what specific labels and markers to use—we recognized and needed to navigate the complex ways in which social scientific, computational, and cultural epistemologies of identity and classification came into tension. In the next section, we canvas some of our internal deliberations over this selection of terms, and the associated politics of categorization and recognition, in the development of the experiment and analysis of its results.

## Misreading and Misrecognition: Working With Intersectionality in the Algorithm

In our selection of terms around disability, our discussions centered on linguistic variances in the way disability identities are described across medical, colloquial, and advocacy or community contexts. As one key example, two of the most commonly accepted versions of language to describe disability are person-first (e.g., people with disability) and identity-first (e.g., a disabled person, a Deaf person). While preferred usage varies across national contexts, organizations, and individuals, both versions respond to the historical and ongoing dehumanization and denial of the personhood of disabled people. Person-first language gives primacy to personhood, while identity-first language affirms disability as a form of belonging to a wider cultural group and community. A fundamental premise of both critical disability scholarship and disability rights and advocacy is that nondisabled people should be led by and respect the choices of individual people with disability to use the language they use about themselves.

Language models are, however, trained on data sets that are weighted toward historical and mainstream language use. This presents one of several ethical dilemmas in algorithmic experimental design: whether to adopt language use that is preferred—but less likely to trigger associations that reveal bias—or anticipated—but which knowingly violate the above premise. This dilemma acknowledges that temporal, social, and political disjunctures exist between the nuanced and evolving politics of identity "on the ground" and linguistic parameters within language model training sets. For term selection for our analysis, we followed the National Center of Disability and Journalism's *Disability and Language Style Guide* on the basis that these recommendations represent a balance between scientific and colloquial language use and are also informed by disability community advocacy. However, these are inherently imperfect methodological compromises that leave invisible particular contestations around categorization and classification that have a bearing on critical disability epistemologies and politics. For example, we used both Deaf and Autistic in language generation prompts to determine and measure bias with the broad category of disability. Yet some Deaf and Autistic communities reject labels of disability altogether. Other conditions like obsessive-compulsive disorder (OCD) and schizophrenia, while categorized as psycho-social disabilities in the National Center of Disability and Journalism's *Disability and Language Style Guide*, are varyingly excluded and included within disability discourse and identities across medicalized, service provider, and individual contexts. The categorization, re-categorization, inclusion, and exclusion of identities across these fields have varying effects in relation to social marginalization. Inclusion into a specific category may, for example, determine access to services while simultaneously producing forms of social stigma.

A further conversation between the team involved the metaphorical use of terms such as disabled, blind, and deaf in English. From the perspective of the computational methodology used to assess intersectional bias in these language models, such metaphorical uses would usually be excluded from analysis due to their capacity to skew the types of sentiment generated by tests by producing negative—or sometimes, positive—sentiment around disability terms even when not attached to social identity. As one example, we noted the term "disabled" itself, when not coupled with a human subject ("person," "transgender person," "woman" or "man"), invariably associates with equipment (e.g., computer hard drives), reflecting the inclusion of a large corpus of technical literature that biases model predictions toward the machinic at the expense of the social. To counter such effects, we amended the construction of prompt experiments to ensure disability (and religion) adjectives are bound to a noun signifying a human (gendered or ungendered) subject. For example, "blind" would need to be accompanied by any of (within the limited set of terms used in our analysis) "transgender person," "person," "woman," or "man." While this addressed the immediate methodological issue by bracketing out some metaphorical uses of terms in our analysis, this adjustment ignores how metaphorical uses of disability signifiers, which most often attach to negative sentiments, are a critical dimension of the systemic social marginalization of people with disabilities (Ben-Moshe, 2005; Davis, 2016). They cumulatively contribute to the production of stigma through language, even when not attached linguistically to "bias" as it is usually defined in language models experiments—that is, as negative sentiment against an individual or social group. Adoption of a critical sociological approach would argue for, then, the inclusion of these metaphorical uses of disability terms. At the same time, this inclusion would also produce results that lack comparability to other tests or skew the measurement of bias against disability identities in different ways from other social categories. This constitutes one limitation of both our study and others that seek similarly to assess machinic bias "fairly": considerations of the power of metaphorical language in producing bias beyond literal descriptors of people with disability cannot be bracketed out of the comprehensive understanding of language use and the production of bias, whether within AI systems or in lived experience. As social and computational artifacts, AI systems conserve outdated, offensive, and biased metaphors from sources ranging from chat forums to historical documents and medical texts. These metaphors may not be made evident in "pure" tests of algorithmic bias, meaning apparently "unbiased" or "de-biased" systems retain stigmatizing metaphors.

Finally, to understand intersectional bias and its relationship with single category bias, we used categories of gender, religion, and disability. This was a choice partially driven by the limited amount of research on language model bias on religion and disability in comparison to race and gender (Magee et al., 2021) but also partially by the methodological difficulty of defining language markers for race, ethnicity, and culture beyond limited racial categories. Neither social nor sociological language has universal meanings in terms of racial and ethnic categorization. Our overall commitment in the project to the locally situated understandings of identity within our research context was muddied by the overwhelmingly U.S.-centric focus within the existing literature on algorithmic bias, which, although critical as foundational texts of the field, can render invisible the "polyvalent mobility of race" (Kunz, 2016) across other contexts. Binary categories like Black/White, for example, signify differing identities and histories globally. As Aboagye (2018, p. 76) writes, in the "black bi-cultural landscape" of Australia, concepts such as "blackness" and "indigeneity" are intersecting ideas that are "distinct to the nature of black Australia and the discursive field of black Australian(ness)." While language model datasets are often empirically positioned as universal technical tools, they slant inevitability toward United States and Euro-centric usage in terms of volume and frequency of language use.

A localized and intersectional lens troubles other categories of race too. For example, the meaning of being categorized as "Asian," whether in relation to demographic data collection or everyday usage, varies across migrant settler contexts globally. In Australia, "Asian" usually denotes East Asian heritage, while in the United Kingdom it attaches predominantly to South Asian communities (Watkins et al., 2017). A generalist demarcation of racial or ethnic categories can still provide evidence of bias in language generation models. However, the question of who is stigmatized by this bias in relation to local and lived experience would likely evade such a generalist view. Furthermore, the geographic concentration of studies of algorithmic bias, which remains heavily centered on North America, produces a second order of bias along national and regional lines, within which the localized ethnic, racial, and disability identities of communities in other national contexts (and in particular in the Global South) are excluded from both empirical and theoretical knowledge around algorithmic bias and "fairness." Despite the volume of training datasets, they remain tethered to specific understandings of social identities and categories that can render specific local identities and forms of marginalization invisible or misrecognized. As Bridges (2021, p. 8) writes in relation to digital identity resolution, the "unfreedom to classify and identify oneself" is central to practices that "refuse" and 'back-talk'—rather than fix—algorithmic and data-driven surveillance of the marginalized.

Language is both a tool and a product of computational attempts to measure and mitigate bias that rely on rendering social identities into specific categories that are recognizable and translatable both algorithmically and socially. Yet, the slippages and stickiness of language materialize within and through methods and turn us to iterative dialogues around the trouble of meaning, power, and classification. Ultimately, these tensions are entrenched within the broader "trouble" of AI systems as an object of sociological inquiry, as systems in which categorization, classification, and labeling are foundational to their function. The computational methods used to provide both evidence and mitigation of their biases are inevitably constrained to "speak the same language" of classification in ways that reproduce the power dynamics inherent within these systems. As we found in our research experiments, misrecognition and misreading in AI systems are potentially mirrored in attempts to challenge their biases. Yet our field experience also brings to light how AI systems like language models exist as continuations of historical systems of social classification and categorization, including sociology as a scholarly discipline, that have always been violent and exclusionary and served the interests of domination. The interdisciplinary methodological dialogue has brought instances of misrecognition to light, but we continue to grapple, at the point where disciplines and methods intersect, with "the reductive violence of naming and categorisation" (Bridges, 2021, p. 3), both on the ground and in the algorithm.

## Conclusion

Although as researchers we were involved in both vignettes, the two differ markedly in methods, data, competency demands, and approaches to analysis. What draws them together? Each dwells upon the role of language (metaphor, prediction) in producing the image of alterity: in the first vignette, humans of a machinic other; in the second, machines of a human other. As researchers mediating those AI encounters, we are subject ourselves to common experiences of misreading and misrecognition too: projections of what we anticipate both human and machinic subjects will articulate. The deliberations in the design of the algorithmic experiment inscribe, in particular, a desire to test the linguistic productions of machines "fairly," by opting as much as possible for what we imagine as "neutral" descriptors of gender, disability, and religion. Conversely, in constructing tests of bias, we can also be seen as projecting onto the machine an expectation that it might be "more than human" in its avoidance of biased and prejudiced associations with different markers of difference and intersectionality—an expectation that, as we show, the machine in its various forms fails to live up to. Indeed, we also note that at least with respect to language, the machine necessarily depends upon human-generated data and human (or human-trained algorithmic) interpretation. Whether desirable or otherwise, it cannot escape the structural constraints of human language use, under which bias is a continuously and socially negotiated feature.

Placing these two methodological vignettes from very different disciplinary traditions in close proximity is important to our aim to think through concepts of intersectionality, algorithm, bias, and metaphor across methods that historically have been clearly aligned to distinct social and technical disciplines and traditions. We do not think such placement succeeds in any naive closure of epistemic fissures between those disciplines nor in any greater alignment of human and machinic imaginaries. Rather, when taken together, what the methodological gaps, concessions, and occasionally unresolvable questions within these research encounters unveil is the salience and complexity of language and metaphor, as temporally, socially, and politically situated means by which intersectionality, social marginalization, and visions of AI, are produced, reproduced, and perhaps also transformed. The first vignette from our qualitative fieldwork illustrates the pedagogical and interpretive limits of speculative thinking within qualitative research encounters. It also highlights the circuitous and iterative layers of knowledge, cultural, and linguistic translation that occur around such moments, as participants explain cultural symbols to researchers and as researchers translate linguistic and cultural meanings between each other during the processes of data collection, transcription, and analysis. It raises the question of how to disentangle the palimpsest of meaning surrounding a speculative metaphor like a care robot that is simultaneously metaphorical and more-than-metaphorical, as it is in the process of "coming into being" as a functional AI architecture that may materially shape the futures of research participants. As Ganesh (2020, p. 3) notes, "artificial intelligence is constructed through a fertile and messy exchange of metaphors about human and machine [which are] powerfully entangled with epistemology even when they are not accurate." We are left, then, to reflect on scholarly and ethical responsibilities to see the research encounter as a space not only to analyze these metaphors but also to consider the import of their dialogical transformation and destabilization within the communities in which we research.

Our second vignette highlights potential misrecognition and misreading in our experiments with language models, revealing the points at which our determinations around method opened challenges to the universality not only of algorithms but of the computational processes of their de-biasing in relation to local contexts, language, and evolving self-determination and identity. As Crawford argues, the politics of classification is "baked into the substrates of AI" in ways that render its mitigation always and inevitably methodologically partial (Corbyn & Crawford, 2021). Yet, the same is true of social inquiry that, even if critical of and attentive to the power dynamics that work through the production of categorization and classification, still relies on modes of classification to determine the boundaries of both social identities and research designs. Such questions are crucial to grapple with if research on bias mitigation in AI systems seeks, as we do, to look beyond technically driven remediating strategies for algorithmic biases and toward engaged approaches to addressing inequity in collaboration with communities. The rights of community members to self-determine their identities across intersectional social differences and within their own social contexts are at stake in the design and implementation of such interdisciplinary research. The emphasis on social power that grounds intersectionality as theory "reminds us that maximising algorithmic fairness does not substitute for addressing historical injustice or protecting the most marginalised" (Bauer & Lizotte, 2021, p. 99) but further reiterates the limits of algorithmic interventions that cannot escape the biases embedded within language itself as a power structure and, in particular, within language that seeks to classify social difference in bounded ways.

A further line of connection between these two seemingly disparate research encounters is how the projections of both the "caring robot" and the "fair algorithm" condition our collective fantasies of AI: fantasies that are shaped and mediated through research encounters as well as through technical, regulatory, and literary discourses and tools. An intersectional lens ultimately demands not merely that such visions be made "inclusive" of intersectional difference but that our methodologies work to foundationally challenge whether "care" and "fairness" are operable goals of AI systems, question the limits of these visions as solutions to entrenched social marginalization, and consider how we collectively realize alternative ways to talk AI into being. Following Bridges's (2021) analysis of the political potentials in "digital failure," we can similarly read the gaps and cracks in our methods as openings through which to begin to unsettle the language of AI, and the meaning of the research encounter as a space to produce and reflect on such unsettling.

These gaps and cracks also point to practical aims, of the kind the wider project that houses these two case studies is seeking to further. Researchers and activists have long pursued ways of including those who are affected by technology into the design and evaluation of these systems. Recent work by researchers working with Syrian refugees on algorithms (Kasapoglu et al., 2021; Masso & Kasapoglu, 2020) argues for a reconsideration of algorithmic governance that would involve more collaborative and flexible feedback loops between various data "agents"—authorities, experts and users—to improve technocratic means of "caring for the self." Arguably work by large technology firms is already at least performing feedback loops through funded initiatives like *AI for Accessibility* (Microsoft), *PAIR* (People + AI Research at Google), and *Responsible AI* (Meta). A challenge for such approaches in AI is simply the scale at which language models are developed and trained—nuanced user feedback, particularly in relation to groups that represent intersectional minorities, has to be integrated into systems that take months and millions of dollars to train. As our first vignette shows, AI can and needs to be thought beyond (often masculinized) races toward "state-of-the-art" results and accompanying prestige. Careful attention to diverse community aspirations and settings—from care and support to leisure—should, we argue, produce greater diversity and eclecticism in what is considered legitimate "AI" research and products. One area for exploration here would be the application of user feedback (Wellner & Rothman, 2020) or feedback loops (Kasapoglu et al., 2021) to practices of fine-tuning machine learning systems with the aim of, if not eliminating forms of bias entirely, at least articulating other less stereotyped images of intersectional subjects.

## Acknowledgement

The authors would like to acknowledge the significant contribution of our research partners to the research activities we reflect on in this paper, in particular Dr Snow Li at Your Side Australia, Rachel Haywood at Western Sydney MRC and Dr Lida Ghahremanlou at Microsoft.

## Declaration of Conflicting Interests

The author(s) declared no potential conflicts of interest with respect to the research, authorship, and/or publication of this article.

## Funding

The author(s) disclosed receipt of the following financial support for the research, authorship, and/or publication of this article: The research has been funded by the Australian Research Council: LP190100099

## ORCID iD

Liam Magee  https://orcid.org/0000-0003-2696-1064

## Author Biographies

**Shanthi Robertson** is a leading researcher and media commentator on contemporary migration, cultural diversity, and urban social inclusion. She has worked in partnership with local government on enhancing social inclusion in multicultural communities and has provided expert policy inputs at a federal level.

**Liam Magee** is a researcher working on social impacts of emerging technology systems, standards, and infrastructure. A former software developer and project manager, as an academic Liam continues to collaborate with private and public sectors on innovative research projects.

**Karen Soldatić** is a leading global disability scholar with 20 years of experience as an international (Cambodia, Sri Lanka, Indonesia), national, and state-based senior policy advisor, researcher, and educator.